# Anti-seizure medication tapering correlates with daytime delta band power reduction in the cortex

Guillermo M. Besné[1], Nathan Evans[1], Mariella Panagiotopoulou[1], Billy Smith[1], Fahmida A Chowdhury[3], Beate Diehl[3], John S Duncan[3], Andrew W McEvoy[3], Anna Miserocchi[3], Jane de Tisi[3], Mathew Walker[3], Peter N. Taylor[1,2,3], Chris Thornton[1], Yujiang Wang[1,2,3]*

1. CNNP Lab (www.cnnp-lab.com), School of Computing, Newcastle University, Newcastle upon Tyne, United Kingdom

2. Faculty of Medical Sciences, Newcastle University, Newcastle upon Tyne, United Kingdom

3. UCL Queen Square Institute of Neurology, Queen Square, London, United Kingdom

* Corresponding author:

Yujiang Wang

CNNP Lab (www.cnnp-lab.com), School of Computing, Newcastle University, Newcastle upon Tyne, United Kingdom

Yujiang.Wang@newcastle.ac.uk

## Short title:

EEG power reduction in ASM tapering

## Abstract

Anti-seizure medications (ASMs) are the primary treatment for epilepsy, yet medication tapering effects have not been investigated in a dose, region, and time-dependent manner, despite their potential impact on research and clinical practice.

We examined over 3000 hours of intracranial EEG recordings in 32 subjects during long-term monitoring, of which 22 underwent concurrent ASM tapering. We estimated ASM plasma levels based on known pharmaco-kinetics of all the major ASM types.

We found an overall decrease in the power of delta band ($\delta$) activity around the period of maximum medication withdrawal in most (80%) subjects, independent of their epilepsy type or medication combination. The degree of withdrawal correlated positively with the magnitude of $\delta$ power decrease. This dose-dependent effect was evident across all recorded cortical regions during daytime; but not in sub-cortical regions, or during night time. We found no evidence of a differential effect in seizure onset, spiking, or pathological brain regions.

The finding of decreased $\delta$ band power during ASM tapering agrees with previous literature. Our observed dose-dependent effect indicates that monitoring ASM levels in cortical regions may be feasible for applications such as medication reminder systems, or closed-loop ASM delivery systems. ASMs are also used in other neurological and psychiatric conditions, making our findings relevant to a general neuroscience and neurology audience.

## Introduction

The potential for Anti-seizure medications (ASMs) to prevent seizures is well documented[1,2]. There is, however, a dearth of evidence for the precise effects of ASMs on the neurophysiology of humans *in vivo*. For example, carbamazepine is thought to slow background EEG activity[3], but this effect may be highly subject-specific[4]. The cellular mechanisms of action also differ widely between ASMs, leading to varying effects of individual medications[3] and poly-therapy effects that are highly complex. Yet, the group of patients requiring poly-therapy are the most challenging to manage clinically - often having uncontrolled seizures, and poor adherence to medication[5]. The mechanisms behind drug-resistance remain unclear and are most likely complex[6]. Hence, active research efforts - many using neurophysiological investigations - often focus on these drug-resistant patients on poly-therapy. However, due to limited opportunity for systematic investigations in humans, the effect of ASMs is poorly understood and rarely accounted for in these studies.

ASM tapering during long-term electroencephalography (EEG) monitoring, performed in preparation for resective epilepsy surgery, provides a rare opportunity to study poly-therapy ASM effects in humans in a controlled manner. Individuals are given ASMs at set times, and controlled tapering of the ASM is often performed to provoke seizures for the identification of the seizure onset zone[7]. This procedure provides an opportunity to study how medication dose relates to the characteristics of the EEG. Previous work indicated that EEG changes can be detected in response to temporary ASM withdrawal in such settings[8–12]. For example, reduced delta power during drug tapering was reported by[11]. However, to-date, the effects of poly-therapy ASM tapering on the interictal intracranial EEG (icEEG) remain under-explored. In particular, it is currently unknown if ASM tapering has a distinctive effect on different brain areas, or if the relationship between ASM dose and EEG band power is modulated by time-of-day.

We hope that by addressing these questions, we contribute towards a better understanding of drug-resistant patients on poly-therapy, and specifically provide future neurophysiological studies in this group of patients with some insights into the effect of the poly-therapy itself. Further, we envisage that our results will aid in the design of studies investigating the mechanisms of poly-therapy withdrawal, and pave the way to translational outputs, such as medication reminder systems or closed-loop ASM delivery systems to help with poor medication adherence in this group of patients.

## Methods

### Subject and study information

This is a retrospective study of individuals with drug-resistant focal epilepsy undergoing icEEG monitoring and drug tapering as part of the clinical pathway to evaluate the possibility of surgery. This study had no influence on any clinical decisions made, including during the monitoring period. Our retrospective and anonymised data were obtained from the National Epilepsy & Neurology Database[13], composed of individuals with focal epilepsy from National Hospital for Neurology and Neurosurgery, and analysed with the approval of

Newcastle University Ethics Committee (42569/2023). Retrospectively, we included any subjects that underwent icEEG, had subsequent epilepsy surgery and associated pre-op and post-op T1w imaging information and icEEG implantation information between 2008 and 2019. Patients for whom 24h-long icEEG recordings were not available at both baseline and during tapering were excluded (more details on these 24h-long recordings are given below).

Patient data confidentiality is ensured through the National Epilepsy & Neurology Database (NENDB)[13], which has undergone ethical (NHS REC) approval. No identifiable patient data is stored on the NENDB, and the access is strictly limited to data defined by the inclusion criteria of the project. Furthermore, each project receives a new pseudonymisation of subjects to prevent reconstruction of details about individuals across projects. Demographic data (e.g. age, sex) is only provided as a group summary statistic here for the same reason.

## Identifying brain regions recorded by intracranial EEG

We used concurrent neuroimaging data to identify which brain regions were being recorded by the icEEG. Our MRI processing followed the same pipeline as our previous works[14,15], and a summary is replicated below.

The implantation CT or MRI is co-registered to the pre-op T1w MRI. All electrode contacts are manually marked in pre-op space. In cases of brain sagging/shift, this was visually corrected for, and any surface ECoG electrode contacts are projected to the exposed cortical surface using the pipeline developed by[16]. All outputs for each step are visually inspected and adjusted where needed (e.g. co-registration).

We assigned icEEG electrodes to one of 128 regions of interest (ROIs) from the Lausanne 'scale60' atlas[17]. We used FreeSurfer to generate volumetric parcellations of each subject's pre-operative magnetic resonance imaging[17,18]. Each electrode contact was assigned to the closest grey matter volumetric region within 5 mm. If the closest grey matter region was more than 5 mm away then the contact was excluded from further analysis. A 5 mm radius was chosen to follow previous processing conventions (e.g.[14,15]) and as our visual inspection of the imaging data suggested that a margin of error should be accounted for within the range of a few millimetres.

To identify which ROIs were considered Seizure Onset Zones (SOZs), we combined the information from the volumetric parcellation and clinical reports: clinical experts labelled each recorded icEEG channel as SOZ or non-SOZ by visual inspection of the patient's seizures on icEEG. With this labelling, and the mapping between electrode contacts and ROIS, we labelled ROIs as SOZ if a single channel assigned to the ROI was labelled as such. We took the same approach to frequently-spiking regions (label generated by visual inspection by clinicians) and later surgically resected regions.

## ASM tapering and plasma concentration estimation

We modelled ASM plasma concentration to estimate, in each individual, a 24 h period of stable ASM levels prior to the tapering procedure, and a 24 h period when ASM concentrations overall were lowest.

The ASM tapering information during intracranial monitoring was stored as a list of medications for each subject, with the intake dosage over time. As in previous work and literature[9], we converted this time-series of intake into a continuous estimation of plasma concentration using known pharmaco-kinetics. We used Equation *1* to estimate plasma concentration after single medication oral intake, with a first-order absorption and elimination. The parameters used in Equation *1* are specific to each ASM and have been obtained/estimated form several sources[19–21] (Supplementary *S2* for further detail). The three components of this equation (left to right) represent i) $C(t)$: the concentration of the medication in systemic circulation, ii) fraction term: the availability of ingested medication considering metabolism, distribution and protein interaction, and iii) bracket term: the elimination (positive term) and absorption (negative term) rate into systemic circulation. The individual effect of each dose - modelled by single dose pharmaco-kinetics - are combined to model the plasma concentration through the entire icEEG recording, yielding a time series of medication concentration $C(t)$ during the entire EMU stay for each patient (see Figure *1*A, orange line). Please see Supplementary *S2* for further details.

This time series C(t) still contains fluctuations with each intake dose, even if the intake was regular (Figure *1*A, orange line). To simplify the model and extract periods of regular medication protocols *vs.* tapering, we therefore applied a 24 h rolling average (Figure *1*A red line).

$$C(t) = \frac{F * D * Ka}{Vd * (Ka - Ke)} * \left(e^{-Ke*t} - e^{-Ka*t}\right)$$

**Equation.1:** Pharmaco-kinetic equation for plasma concentration of a single dose oral intake with first-order absorption and elimination. C(t), plasma concentration. F, bio-availability. D, dose at intake. Ka, absorption constant. Ke, elimination constant. t, time since intake. Vd, volume distribution.

To normalise the effects of medication across individuals, the ASM level is normalised as a ratio of the 24-h rolling average level, ranging from 1 (regular ASM plasma concentration) to 0 (zero ASM plasma concentration). Many individuals were submitted to poly-therapy (multiple ASMs), in some cases with slightly different temporal tapering profiles for each ASM and reaching different levels or reduction (Figure *1*B red lines). In these cases, normalised levels of all ASMs on a single subject are averaged (Figure *1*B black line). We use this to approximate the average effect of medication tapering. There were also non-tapered ASMs as well as rescue medication (ASMs outside the regular prescription of the individual used to accelerate recovery from tapering and severe seizures). The influence of these medications on ASM levels is included, but they were not considered tapered medications.

To determine the effect of tapering on icEEG features, we identified two time periods of interest to analyse: the regular ASM period pre-tapering (baseline-ASM), which we took as the last 24 hours before tapering (Figure 1B, hatched area); and lowest ASM period (reduced-ASM), 24 hours centred around the point of lowest estimated average ASM levels (Figure 1B, dotted area).

Finally, to estimate an overall ASM level at the point of strongest tapering for each subject, we obtain the lowest point of the overall ASM level (lowest point on thick black line in Figure 1B, marked by black circle).

### EEG data and processing

Our icEEG processing followed the same pipeline as our previous work[14,15], and a summary is replicated below.

Firstly, we divided each subject's icEEG data into 30 s non-overlapping, consecutive time segments. All channels in each time segment were re-referenced to a common average reference. In each time segment, we excluded any noisy channels from the computed common average. Noisy channels were detected automatically using outlier detection across channels on the amplitude ranges ($\pm 3$ standard deviations on z-score data) of the EEG data. Missing data were not tolerated in any time segment and denoted as missing for the downstream analyses. All 30 s segments containing seizure data and other provocation procedures - such as cortical stimulation and sleep deprivation/alteration - were removed and considered missing for our analysis.

To remove power line noise, each time segment was notch filtered at 50 Hz. Segments were band-pass filtered from $0.5 - 80$ Hz using a $4^{th}$ order zero-phase Butterworth filter (second order forward and backward filter applied) to preserve the information on all canonical frequency bands: Delta (1-4 Hz), Theta (4-8 Hz), Alpha (8-13 Hz), Beta (13-30 Hz) and Gamma (30-47.5 Hz, 52.5-57.5 Hz, 62.5-77.5 Hz). To lower data density and preserve the information on faster frequency bands, all channels were down-sampled to 200 Hz.

We then calculated the icEEG band power in the delta frequency band for all channels and each 30 s time segment using Welch's method with 3 s non-overlapping windows. In detail, for each channel in every 2 s window, we calculated the power spectral density and used Simpson's rule to obtain the band power values, which were then averaged over all time windows within a 30 s segment to get the final band power values. We $log_{10}$-transformed the band power values, thus obtaining, for each subject, one matrix of log($\delta$ band power) at the ROI level. The size of each subject's matrix was number of ROIs by number of 30 s segments.

Finally, from this matrix we retained data for the two time periods of interest: 24 h baseline-ASM and 24 h reduced-ASM. Only individuals with proper information on these matrices on at least 1 ROI and 1 common hour on both 24 h periods (e.i. 2pm on both baseline-ASM and reduced-ASM) are included on this study.

We chose to focus on the EEG delta frequency band for simplicity. Similar results are shown in Supplementary *S3* for other frequency bands, but with weaker effects.

## Accounting for time-of-day effects

The time-of-day for the 24 h baseline-ASM period may not be aligned with the 24 h reduced-ASM period. For example, the first hour of baseline-ASM may fall on 14:00, whereas the first hour of reduced-ASM might be 17:00. $\delta$ band power is known to fluctuate strongly on a circadian timescale[15,22], driven by sleep/wake cycles. As we lacked sleep/wake information in our data, we devised a method to account for the effect of the time-of-day: First, band power for each ROI was averaged for each hour of the day. For example, all data collected between 04:00 to 04:59 was averaged to a single value per ROI. If the hour segment had more than 75% of the data missing, the entire hour is considered as missing. Next, data were rearranged to match the time-of-day series from 00:00 to 23:00. For example, if the initial interval corresponds to 04:00, the final 4 hours after midnight were moved to the beginning, resulting in 24 hourly averages beginning at 00:00-00:59 and ending at 23:00-23:59. This protocol will match stages of circadian fluctuations - such as sleep/wake cycles - between the defined ASM periods to factor these out and reveal potential differences in terms of ASM reductions. Matrices in Figure *2*A&B demonstrate the hourly average and alignment procedure for an example subject's baseline-ASM and reduced-ASM periods respectively, where each column corresponds to an hour, and each row an ROI.

The difference between the two periods was obtained by subtracting the baseline-ASM matrix from the reduced-ASM matrix to produce Figure *2*C. This resultant difference matrix captures the effect of ASM tapering on an hourly basis, accounting for the time-of-day effect. Finally, to achieve an average estimate of the effect of ASM tapering, we average the difference matrix over time (Figure *2*D), thus obtaining a vector of tapering effects for each region recorded.

## Statistical analyses

We provide p-values as an additional reference to reported effect size, but we do not threshold the p-values to dichotomise our data based on statistical significance further or make subsequent analysis decisions.

To determine if there were any differences between the baseline-ASM and reduced-ASM periods across subjects, we obtained the time-averaged band power difference between them in each region and each subject. We tested if the distribution of band power differences between baseline-ASM and reduced-ASM is different to 0 using non-parametric Wilcoxon signed rank tests. We subsequently correlated the magnitude of the ASM tapering with the band power change between baseline-ASM and reduced-ASM using Spearman's rank correlation. To determine regional effects on the band power changes, we divided all ROIs into two categories: cortical and sub-cortical regions (in this case, Amygdala and Hippocampus); and we repeated the analysis for each category.

An additional, confirmatory analysis using hierarchical models accounting for subject-level differences in the location and number of regions recorded is presented in

Supplementary *S4*. For simplicity, we only present the simpler, single-level model in the main text. These hierarchical models are used to confirm the relationships with ASM tapering magnitude, the regional difference between sub-cortical and cortical regions, and to investigate the effect of cortical lobes. Additionally, hierarchical models were used to analyse some confounding factors: differential effects of seizure onset zones, frequently-spiking regions, and later surgically removed regions; tapering of different ASM types; and the influence of recovery from anaesthesia post implantation.

Finally, we also disaggregate for the time-of-day effect. Instead of averaging over the 24 h period to study the band power change, we repeated our analysis for each hourly average (i.e. on the matrix of Figure *2*C). Here we opted to show the effect size of band power reduction and the associated confidence intervals for each hour. We did not perform further statistical tests to compare any subgroups due to limited sample size, and these results are therefore to be interpreted as indicative.

### Data Availability

Anonymised EEG band power data and ASM intake schedule, along with analysis code will be available on Github: *https://github.com/cnnp-lab/2024_ASM_EEG*.

## Results

### Subject and data characteristics

We included 22 individuals who underwent ASM tapering and had two 24 h icEEG data segments available for baseline-ASM and reduced-ASM states. The male:female distribution was 8:14, with an average age of 31.3(±8.6) years. In terms of epilepsy, individuals were diagnosed with temporal(16) and frontal(6) lobe epilepsy, affecting either left(10) or right(12) hemispheres.

We further summarised the ASM types used among the 22 individuals in Supplementary *S1*. The most frequently-used ASMs were Levetiracetam (45%), Lamotrigine (41%) and Clobazam (36%). The average time between the start of medication tapering and reaching the minimum ASM level was approximately 2 days and 10 h (Supplementary *S1*).

In addition, we analysed 10 individuals with icEEG monitoring, but no ASM tapering in Supplementary *S4.5* to support our main results.

### icEEG band power decreases are correlated with medication tapering

We found an overall decrease in $\delta$ band power across all subjects between baseline-ASM and reduced-ASM states, after accounting for time-of-day effects. The distribution of $\delta$ band power changes over all regions and subjects (Figure *3*A) and is, on average, decreasing (signed rank test effect size: -0.489, p<0.001). Averaging over all regions first in each subject yields similar results (signed rank test effect size: -0.626, p=0.003). As reported in Supplementary *S3*, a decreased band power was also found in other frequency bands.

However, the lowest average value was seen in the $\delta$ band, and we therefore focus on this band for the main results.

Considering that each subject has been tapered off their ASMs to a different degree, we investigated if the $\delta$ band power decrease was tapering dose-dependent. Figure *3*B shows the $\delta$ band power decrease against the relative percentage of ASM tapering for all subjects and ROIs. We observe a tendency of greater decrease in $\delta$ band power as ASMs are tapered to a lower level (Spearman's rank correlation coefficient: 0.535, p<0.0001, confidence interval: [0.458 0.604]). Supplementary *S4.1* additionally shows confirmatory results using a hierarchical model to account for regional and subject-level effects (minimum ASM level estimate: 1.377; p: 0.008; confidence interval: [0.352, 2.401]).

We further investigated ASM-type-dependency by classifying medications based on their physiological target (Supplementary *S4.3* for further details). Using a hierarchical model we observed that no ASM class had a distinctive effect on $\delta$ band power. We report all relevant statistics in Supplementary *S4.3*.

We also considered if any additional long-term effects might be present, such as the wearing-off of anaesthesia following surgery for implantation. We therefore investigated if the time between surgery and the baseline-ASM period impacted our results, and found no evidence of this (see Supplementary *S4.4* for full statistics tables).

Finally, to further test the robustness of the observed dose-dependent effect of $\delta$ band power decrease, we added the 10 subjects that were not ASM-tapered to our analysis as additional data points at 'full-dose'. These additional subjects did not change the model coefficients substantially (and all p-values remained similar), further supporting the robustness of the model. Full details can be found in Supplementary *S4.5*.

## Differential regional effects of ASM tapering

We disaggregated the previous analysis into cortical and sub-cortical (Amygdala and Hippocampus) regions, and found that the $\delta$ band power decrease was primarily seen in cortical brain regions (Figure *4*).

In the sub-cortical regions, we found no evidence of $\delta$ band power decreases (signed rank test effect size: -0.076, p=0.628), whilst in the cortical regions the previously reported decrease remains (signed rank test effect size: -0.535, p<0.001). The relative percentage of ASM tapering influences both cortical and subcortical regions, although the effect was stronger in cortical regions (Spearman's rank for cortical correlation coefficient: 0.566, p<0.001, confidence interval: [0.488 0.635] and sub-cortical correlation coefficient: 0.362, p=0.021, confidence interval: [0.057 0.605]). Further analysis using a hierarchical model confirmed the dose-dependent effect on cortical regions but showed no evidence of dose-dependent effect for sub-cortical regions (see Supplementary *S4.1*). We used this same modelling approach to investigate whether specific cortical lobes are driving the observed effect, but found no evidence of this (Supplementary *S4.1*).

Finally, we tested if seizure onset regions showed a differential effect compared to other regions, factoring in the observed cortical and sub-cortical difference and subject-level effects, but we found no evidence of this (see Supplementary *S4.2*). Similarly, regions that were marked as frequently-spiking, and regions that were later removed by epilepsy surgery showed no differential effect (Supplementary *S4.2*).

## Circadian profile of ASM tapering effect

In all prior analysis, we analysed the average effect of ASM tapering over a 24 h period, after accounting for time-of-day effects by calculating band power differences on matched hours between baseline-ASM and reduced-ASM. In our final analysis, we disaggregated our results by the time-of-day, on an hourly basis. Figure *5* shows a stronger decrease of $\delta$ band power during the daytime, particularly in cortical regions (Figure *5*).

## Discussion

In this study, we confirmed previous observations that ASM tapering is associated with neurophysiological changes in icEEG[23]. More specifically, we saw a reduction in band power, particularly in the $\delta$ band, in all subjects and primarily in cortical regions. Importantly, the observed reduction was linearly dependent on the strength of the tapering, further supporting a direct causal link.

A variety of ASM-mediated neurophysiological effects - including the effects of tapering and withdrawal - have been reported[3,12,24,25], and our study is most comparable in its retrospective, icEEG based design based to that of Zaveri et al[11]. We find it encouraging that we were able to directly reproduce their result of band power reduction, which was most strongly expressed in the $\delta$ band, followed by a weaker reduction in $\theta$ and $\alpha$ bands, with negligible effects in $\beta$ and $\gamma$ bands. We further replicated the "wide-spread" nature of this effect as reported by Zaveri et al[11]. This independent replication in a completely separate cohort, in a different country, suggests that this is a reliable effect to leverage for future translational research.

We substantially expanded on previous studies by additionally presenting a dose-dependent effect, a differential effect in cortical regions, and by considering time-of-day. We were able to show that the dose-dependent $\delta$ band power decrease with tapering is present only in the cortex. The Amygdala and Hippocampal regions did not see a substantial or systematic change in band power with ASM tapering. This observation of a cortical *vs.* sub-cortical effect suggests that the observed band power change is not a purely technical consequence of e.g. drift in electrode properties[26,27], which we further confirmed with a separate cohort that were not tapered in their ASMs (Suppl. *4.5*).

It is also worth clarifying that the reduction in $\delta$ band power does not represent "slowing" or "speeding up" in the traditional sense, as we see no evidence that the power of a particular band is migrating to a neighbouring band. Rather our observation is that almost all frequency bands see a reduction, but it is strongest in the $\delta$ band. At this point, we can only speculate on the interpretation and mechanism of a dose-dependent power decrease across the cortex with lowered ASMs in this patient group. It may indicate change in local or global synchronisation of neurons, or the return of some cognitive functions suppressed by ASMs, which is a well-known side-effect[28]. However, we believe that, before jumping to conclusions, many future studies are required, for example, we need to understand on an individual level, what the drug-naive baseline levels are relative to the full-dose and tapered dose. We also need to establish if there is a differential effect in drug-resistant patients. Finally, duration of exposure to ASMs may also play a role.

Our sample size was too small to carry out a region-specific analysis, but our analysis did not show evidence of further spatial differentiation within the cortex. We also saw no differential effect in seizure onset (SOZ), frequently-spiking, or later resected regions. This is in contrast to some previous studies that found a differential effect in other signal metrics in the SOZ[8,29]. However, the effects reported by Sathyanarayana et al.[8] were within higher frequency bands and used different EEG signal metrics (e.g. signal entropy). Interestingly, there are also studies that report no difference between SOZ and non-SOZ regions. Indeed, Meisel *et al.* show a near 1:1 correlation between SOZ and non-SOZ regions[12] in terms of a measure of EEG signal synchrony during interictal periods. The authors concluded that all regions may be equally suited to detect global brain synchrony changes with ASM tapering. We hence conclude that different EEG signal metrics/markers are most likely differentially sensitive to SOZ with ASM tapering. In our case, $\delta$ band power in the cortex does not appear to be sensitive to the changes associated with SOZ.

Studying the specific effects of ASM types was also not possible, as no subjects were on exactly the same combination of ASM types and baseline dosages. The ASM classification aided, and allowed us to perform, hierarchical modelling to observe possible changes. The fact that none of the defined classes had a significant effect can be explained by the heterogeneous effects of ASM types within the same class[24]. However, there was a trend in multi-target ASMs, which could indicate a predominant effect of Sodium Valproate, which is known for having a strong effect on $\delta$ band power[30].

Together, these observations may suggest that future work investigating the mechanisms of ASMs in epilepsy needs to account for the major structural and functional differences in the Amygdala and Hippocampus compared to the cortex, but specific epileptogenic alterations in the brain may be too heterogeneous to study in this context and much larger homogeneous samples are needed.

Our final key observation was that the tapering induced $\delta$ band power decrease was also dependent on the time-of-day, with the daytime hours showing the largest decrease. Unfortunately, information about the sleep/wake status of each subject was not available, hence we can only assume that the daytime hours primarily contain awake-states across the cohort[31]. With this assumption, our results could be interpreted as a $\delta$ band power decrease during wakefulness as opposed to sleep. Sleep and drowsiness have their own neurophysiological architecture and markers[32], and epileptic spikes occur more frequently in sleep. We therefore cautiously hypothesise that ASM tapering does not have a strong or consistent effect on band power composition, or spike properties during sleep. The latter hypothesis is additionally supported by our observation that the frequently-spiking regions did not show evidence of a differential effect compared to other regions (Suppl. *4.2*). Naturally, these hypotheses need to be looked at in isolation and tested further in terms of ASM type, sleep architecture, and spike neurophysiology.

Despite clear advantages in our study, including dose-dependency analyses, regional differentiation, and accounting for time-of-day effects, our relatively small sample size limited our statistical power to discover potential effects in combinations of factors (such as type of epilepsy, or the combination of ASM types and brain regions). Furthermore, our ASM plasma levels were estimated based on pharmaco-kinetic assumptions, and not validated with actual plasma measurements. We also did not have information on the sleep/wake state of the subjects, limiting us to a time-of-day analysis only. Finally, our results should only be interpreted in the context of a temporary withdrawal (tapering) of

ASM from steady-state levels in an adult, drug-refractory, and "difficult-to-treat" cohort; they may not be applicable in the context of adjusting medication dose for individuals starting on ASMs.

Concretely, we would welcome the following future studies: (i) Independent replication in a larger icEEG cohort with actual ASM plasma levels measured, where a cross-validated performance of the biomarker ($\delta$ power decrease, or others) can be obtained. Furthermore, effects of aetiology, ASM types, sleep, etc. could be studied better in a larger sample. (ii) Given that icEEG is costly and invasive, we would welcome independent replication in less invasive modalities (scalp EEG or MEG/OPM). Given our results, these studies could initially focus on the cortex and short segments of awake resting state across different levels of ASMs. (iii) Studies on a shorter time scale (e.g. continuously tracking ASM plasma levels and EEG signals on an hourly or half-hourly basis) could extract relevant signal features that best track ASM level changes for more immediate closed-loop application. Our own study, and others, may serve as a starting point. (iv) To understand the interplay of chronobiology (and time scales larger than 24h), epilepsy, and ASMs, ultra-longterm recordings would also be crucial. Ideally all studies would include some concurrent measures of cognitive performance, such as reaction time; and several candidate EEG signal measures should be evaluated and compared to be able to answer questions about mechanisms.

With this study, we hope to provide key insights toward both biomarker discovery for temporary ASM withdrawal[8], and inform future neurophysiological studies in drug-resistant focal epilepsy. If validated in e.g. scalp EEG, or subcutaneous EEG, this approach offers a non-invasive or minimally invasive way to assess medication effects in real time. It could enable medication reminder systems[33,34], and closed-loop ASM delivery systems[35] as new treatment strategies. Using an EEG marker has the clear advantage that we are measuring the medication effect in the central nervous system. ASMs are also used in other neurological and psychiatric conditions, making our findings relevant to application in, for example, anxiety disorders, migraine, neuropathic pain, and depression and mood disorders.

## Acknowledgments

We thank members of the Computational Neurology, Neuroscience & Psychiatry Lab (www.cnnp-lab.com) for discussions on the analysis and manuscript.

We further thank Charlotte McLaughlin and all the NHNN Telemetry Unit staff, and are grateful to all the people with epilepsy whose intracranial EEG data enabled this research.

## Funding

P.N.T. and Y.W. are both supported by UKRI Future Leaders Fellowships (MR/T04294X/1, MR/V026569/1). JSD, JdT are supported by the NIHR UCLH/UCL Biomedical Research Centre.

## Competing interests

The authors report no competing interests.

**Figure captions:**

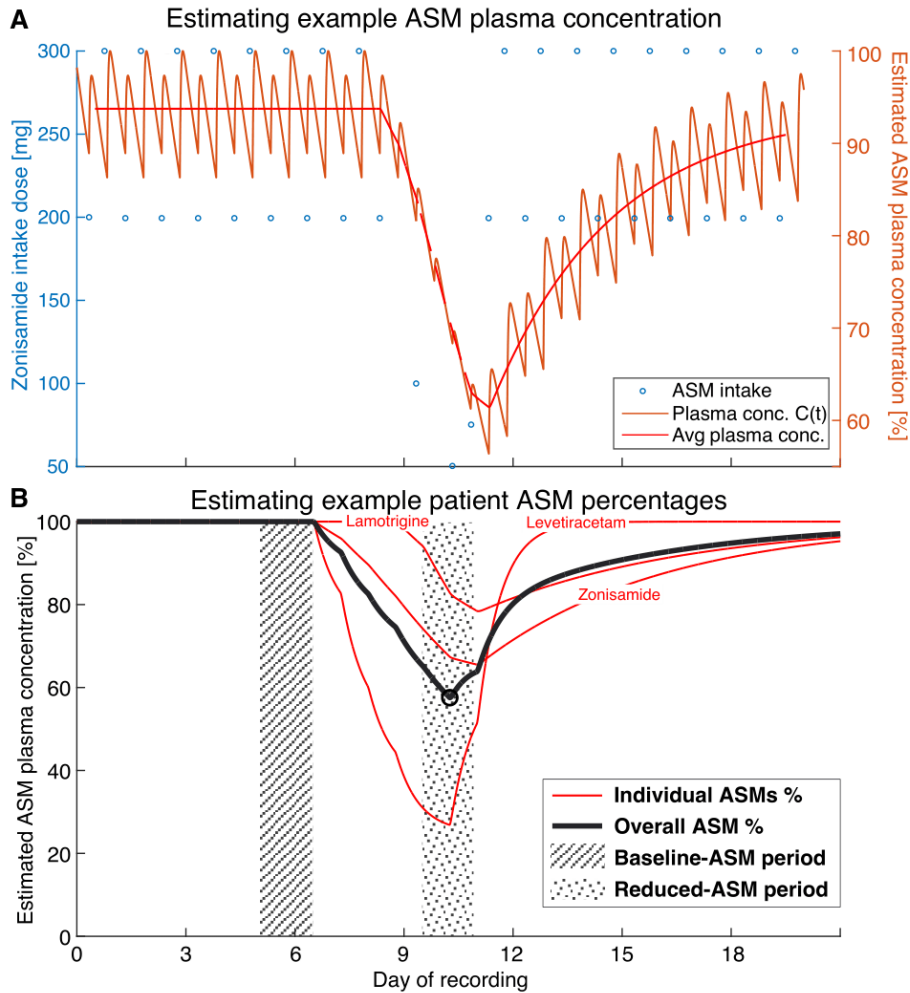

*Figure 1: Estimating Anti-Seizure Medication (ASM) plasma concentration and identifying baseline- and reduced-ASM periods. (A) Example medication (Zonisamide) in an example subject, with intake dosage (blue circles) over time, and the estimated ASM plasma concentration percentage based on the pharmaco-kinetic model (orange line). 24 h rolling mean of the plasma concentration over time is also shown in red. (B) Same example subject as A, with three different 24 h rolling mean ASM percentages (red lines) and the average thereof (thick black line) as an estimate of overall ASM tapering. The baseline-ASM period (hatched area) is identified as a 24 h period before the first drop in the black line and the reduced-ASM period (dotted area) is a 24 h period around the lowest point of the black line (marked by a black circle).*

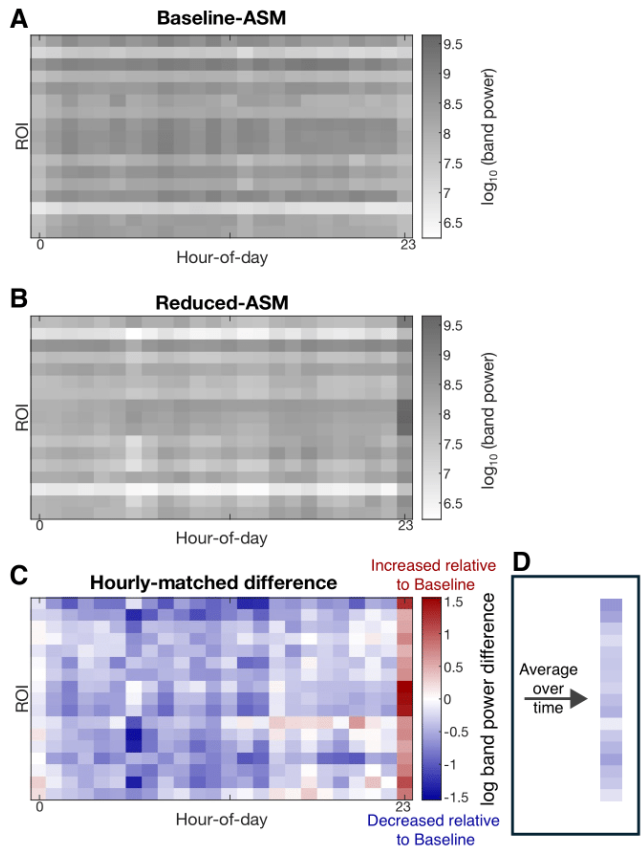

*Figure 2: Log delta(δ) band power difference estimation process for a single subject (n=1) accounting for time-of-day effects. (A) Baseline-ASM band power data matrix and (B) Reduced-ASM band power data matrix, where each row represents a Region Of Interest (ROI), each column an hour of the day and the grey-scale the log δ band power. (C) Colour-coded difference matrix between reduced-ASM relative to baseline-ASM, where red represents an increase in relative band power, blue a decrease and white no change. (D) Average of (C) over time (one value per ROI) following the same colour-code.*

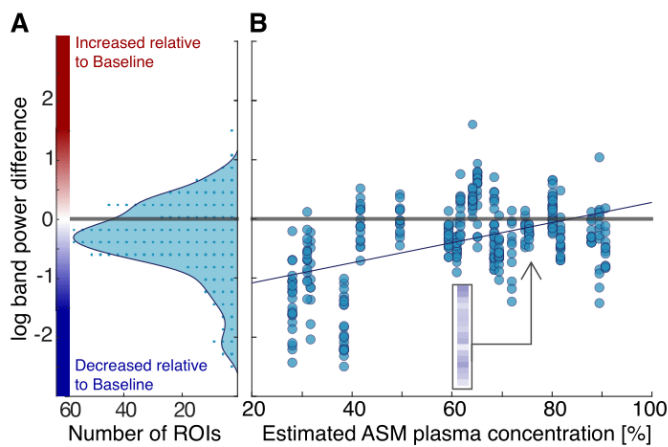

*Figure 3: Tapering effects on log(delta (δ) band power) for all subjects and Regions Of Interest (ROIs). (A) Histogram across all ROIs (N=372) and subjects (N=22) of log(δ band power) difference relative to baseline (each dot equates to 3 data points). Signed rank test revealed an overall statistical decrease (effect size: -0.626, p=0.003). (B) Scatter plot between log(δ band power) difference relative to baseline (y-axis) and the modelled Anti-Seizure Medication (ASM) percentage at the point of strongest tapering (black circle in Figure 1B) in each subject (x-axis). N=372 data points are obtained from 22 subjects, with a variable number of data points per subject due to implantation differences between subjects. Spearman's rank correlation revealed a positive trend between them (coefficient: 0.535, p<0.0001). Insert shows data for all ROIs in a single subject (same as Figure 2D), which in the scatter points all line up on one value of ASM tapering percentage for that subject.*

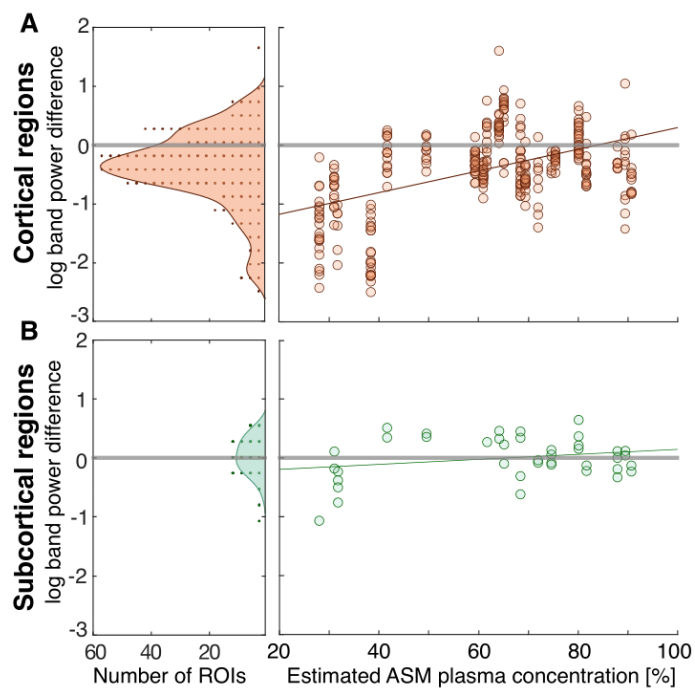

*Figure 4: Anti-Seizure Medication (ASM) tapering reduces delta (δ) band power in cortical but not sub-cortical regions. Plotting convention is the same as Figure 3. (A) The difference between the log(δ band power) before and during tapering in cortical regions (N=332) for each subject (N=22) is shown as a histogram. Signed rank test revealed an overall statistical decrease (effect size: -0.535, p<0.001) and Spearman's rank correlation revealed a positive correlation with ASM concentration (coefficient: 0.566, p<0.001) (B) The difference between the log(δ band power) before and during tapering in sub-cortical regions (Amygdala and Hippocampus) shown for N=40 samples in 17 subjects. Signed rank test did not show an overall decrease (effect size: -0.076, p=0.628) and Spearman's rank correlation showed weaker correlation with ASM concentration than cortical regions (coefficient: 0.362, p=0.021).*

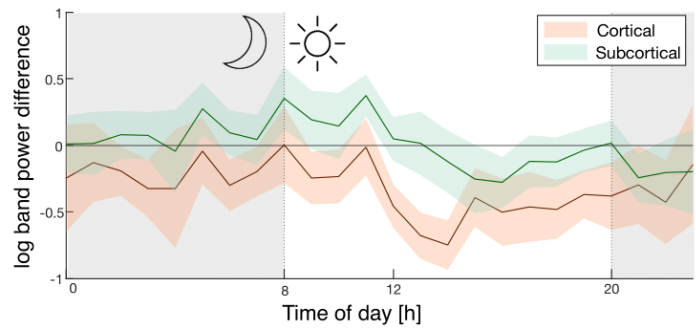

*Figure 5: Log(delta (δ) band power) difference between baseline-ASM and reduced-ASM resolved along the hours of the day. Solid lines represent the mean across Regions of Interest and subjects (N=22), and shaded area represent the 95% confidence interval across subjects, for cortical (orange, N=332) and sub-cortical (green, N=40) regions.*